\documentstyle[12pt]{article}

\begin{document}

\begin{flushright}
Preprint CAMTP/95-6\\
October 1995\\
\end{flushright}

\begin{center}
\large
{\bf Sensitivity of the eigenfunctions and the level curvature distribution in
quantum billiards}
\\
\vspace{0.2in}
\normalsize
Baowen Li\footnote{e-mail Baowen.Li@UNI-MB.SI}
and Marko Robnik\footnote{e-mail Robnik@UNI-MB.SI}\\
\vspace{0.2in}
Center for Applied Mathematics and Theoretical Physics,\\
University of Maribor, Krekova 2, SLO-62000 Maribor, Slovenia\\
\end{center}
\vspace{0.2in}
\normalsize
{\bf Abstract.}
In searching for the manifestations of sensitivity of the
eigenfunctions in quantum billiards (with Dirichlet boundary conditions) with
respect to the boundary data (the normal derivative) we have performed instead
various numerical tests for the Robnik billiard (quadratic conformal map of the
unit disk) for 600 shape parameter values, where we look  at the sensitivity of
the energy levels with respect to the shape parameter. We show the energy level
flow diagrams for three stretches of fifty consecutive (odd) eigenstates each
with index 1,000 to 2,000. In particular, we have calculated the (unfolded and
normalized) level curvature distribution and found that it continuously changes
from a delta distribution for the integrable case (circle) to  a broad
distribution in the classically ergodic regime. For some shape parameters the
agreement with the GOE von Oppen formula is very good, whereas we have also
cases where the deviation from GOE is significant and of physical origin. In
the intermediate case of mixed classical dynamics we have a semiclassical
formula in the spirit of the Berry-Robnik (1984) surmise. Here the agreement
with theory is not good, partially due to the  localization phenomena which are
expected to disappear in the semiclassical limit. We stress that even for
classically ergodic systems there is no global universality for the curvature
distribution, not even in the semiclassical limit.
\\\\
Submitted to {\bf Journal of Physics A}

\normalsize
\vspace{0.3in}
\newpage

\section{Introduction}
The main motive of our present work is to understand and analyze the notion of
{\em sensitivity of the eigenfunctions} of classically chaotic quantal
Hamiltonian systems with respect to boundary conditions, and/or boundary data
and/or the system parameter. It is expected that such sensitivity would
indeed correlate with classical chaos whereas in the classically integrable
systems it would be lacking. It was the important pioneering idea by Percival
(1973), based on the semiclassical thinking, that proposed to classify the
eigenstates and energy levels in regular and irregular depending on whether
they are associated with (supported by) classical regular regions (invariant
tori) or by classical chaotic regions in the classical phase space. This
picture is in fact the basis of the Berry-Robnik (1984) approach to describe
the statistical properties of energy spectra in the transition region between
integrability and full chaos (ergodicity), which has been recently fully
confirmed (Prosen and Robnik 1994a,b). (The role of localization phenomena on
chaotic components implying the fractional power law level repulsion and a
Brody-like behaviour has been understood and demonstrated as well,
showing that in the
strict semiclassical limit, where the localization disappears, the statistics
converges indeed to Berry-Robnik (1984).) Moreover, we have recently performed
the first {\em dynamical} separation of regular and irregular levels and
eigenstates (Li and Robnik 1994c, 1995a,b), thereby explicitly verifying the
ingredients in the Berry-Robnik picture, in particular the Principle of Uniform
Semiclassical Condensation (PUSC) (Li and Robnik 1994a, Robnik 1988, 1993).
\\\\
It was already Percival's idea (1973) to look at the second differences of the
energy levels with respect to some system parameter as an indication of strong
sensitivity (Pomphrey 1974), which is the precursor of the level curvature
concept (Gaspard {\em et al} 1989, 1990).
Of course, in an {\em unfolded} (see e.g. Bohigas 1991) energy level
flow we have stationarity
in the mean so that not only the mean "velocity" but also the
average level curvature is zero, the important point being that the curvature
distribution becomes broader and in fact has some {\em universal
features} in the tail as predicted by Gaspard {\em et al} (1989,1990).
Even at all values of the curvature, after an appropriate normalization
of the curvature, this distribution for classical ergodic systems can
- but need not - be close
to the prediction of the random matrix theory (RMT)
whose functional form has been
conjectured by Zakrzewski and Delande (1993) and proven by von Oppen
(1994,1995). Thus the curvature of individual levels is of course not
indicative of chaos at all but it is its distribution for the ensemble
of levels which matters. We should emphasize, however, that for reasons
explained in section 3 (noninvariance with respect to re-parametrization)
globally the level curvature distribution (after unfolding and after
normalization) {\em need not} be universal even for the class of fully chaotic
(ergodic) dynamical systems. Therefore, the most important
aspect here is just the broadness of the curvature distribution whose
dispersion is of order unity.   In the opposite
extreme of integrable systems we would find just delta spike distribution for
level curvatures, simply because the levels do not interact at all
(degeneracies and level crossings are allowed) so that locally after unfolding
the level flow is just straight lines.
In this paper we address also the intermediate case of mixed classical dynamics
and show that upon the assumptions involved in the Berry-Robnik picture
mentioned before  it can be modelled very easily, but its statistically
significant numerical demonstration is quite hard to achieve,
for which the present work is the first step in this direction.
\\\\
When starting this project we were thinking quite generally about the notion of
sensitivity of eigenstates introduced above and tried to implement some of
these ideas also numerically in the sense of looking at the norm of the second
derivatives of the wavefunctions, but we had to abandon this analysis simply
because it is numerically too massive even for the lowest states. Therefore as
a partial substitute we looked very carefully and in detail at the
energy level flow and the curvature distribution. In section 2 we define the
system and the technique and show the energy level flow. In section 3 we
perform the analysis of the curvature distribution in the regime
of full chaos (ergodicity),
in section 4 we study the curvature distribution in the intermediate case
of mixed classical dynamics, and in section 5 we conclude and discuss
the main results.

\section{The billiard system, technique and energy level flow}

Our billiard system that we use to study the sensitivity of the eigenstates
is defined as the quadratic (complex) conformal map
$w = z + \lambda z^2$ from the unit disk $|z|\le 1$ from the $z$ plane onto the
$w$ complex plane. The system has been introduced by Robnik (1983) and
further studied by Hayli {\em et al} (1987), Frisk (1990), Bruus and Stone
(1994) and Stone and Bruus (1993a,b) for various parameter values $\lambda$.
Since the billiard (usually called Robnik billiard) has analytic boundary it
goes continuously from integrable case (circle, $\lambda=0$) through a KAM-like
regime of small $\lambda\le 1/4$ with mixed classical dynamics, becomes
nonconvex at $\lambda=1/4$ (the bounce map becomes discontinuous), where the
Lazutkin caustics (invariant tori) are destroyed giving way to potential
ergodicity. As
shown by Robnik (1983) the classical dynamics at these values of $\lambda$ is
predominantly chaotic (almost ergodic), although Hayli {\em et al} (1987) have
shown that there are still some stable periodic orbits surrounded by very tiny
stability islands up to $\lambda=0.2791$. At larger $\lambda$ we have
reason and numerical evidence (Li and Robnik 1994b) to expect that the
dynamics can be ergodic. It has been recently rigorously proven by Markarian
(1993) that for $\lambda =1/2$ (cardioid billiard) the system is indeed
ergodic, mixing and K. This was a further motivation to study the cardioid
billiard classically, semiclassically and quantanlly by several groups e.g. by
B\"acker {\em et al} (1994) and B\"acker (1995), Bruus and Whelan (1995).
\\\\
The numerical technique to solve the quantum billiard at various values
of the shape parameter $\lambda\in[0,1/2]$ is the conformal mapping
diagonalization technique devised by Robnik (1984) and further improved by
Prosen and Robnik (1993,1994b) with which, using Convex C3860 machine,
we can at best obtain about 12,000 good levels (in the sense of having an
accuracy of at least one percent of the mean level spacing) at each $\lambda$
and symmetry (parity) class.
Our calculations have been done only for odd parity states for about 600 values
of $\lambda$ such that the lowest 3,000 states have an accuracy of at least
$10^{-3}$ of the mean level spacing. (For technical remarks see (Prosen and
Robnik 1993); in the present case the dimensionality of the matrices that we
diagonalized was always 11,125.) In order to collect all these 600 times 3,000
energy levels we have used 75 days of CPU time.
\\\\
In figures 1-3 we show the energy level flow for fifty consecutive eigenstates
for the entire interval of $\lambda\in[0,1/2]$, evaluated (numerically
calculated) at 600 values of $\lambda$ which is sufficiently dense to make the
appearance of the flow seemingly continuous. We must emphasize that the plotted
energy $E$  here is in fact the {\em unfolded} energy after solving the
Helmholtz equation $\Delta \psi + \tilde{E} \psi =0$
with Dirichlet boundary condition
(vanishing $\psi$ on the boundary $|z| =1$) for the wavefunction $\psi$
corresponding to the pre-unfolded energy $\tilde{E}$, by
using the well known Weyl formula with perimeter, curvature and corner
corrections, given e.g. in Prosen and Robnik (1993) equation (18).
In figure 1 we show states from 1,000 to 1,050, then in figure 2 the
eigenstates from 1,500 to 1,550 and in figure 3 the energy levels from 2,000 to
2,050. In each of these cases we also show some fine structure of the energy
spectrum in two magnified insets.  The main observation is of course that due
to
unfolding the level flow is indeed stationary, that is the mean "velocity"
vanishes, i.e. $<dE_n/d\lambda>=0$. At sufficiently small $\lambda$
like e.g. $\lambda\le 0.1$ the level flow is laminar and all levels are very
flat in this KAM-like regime. This region is then followed by intermediate
values of $\lambda$ between 0.15 and 0.43 where classically we have quite
strong
chaos implying strong level repulsion giving rise to the turbulent level flow
exhibiting many avoided crossings. Finally, at large $\lambda$, say $\lambda
\ge 0.43$, we have classically very strong chaos (in fact ergodicity, mixing
and
K property) which paradoxically gives rise to such a strong level repulsion
that we again find laminar flow. Unlike in the nearly integrable regime of
small $\lambda$ here with increasing energy we would eventually recover the
fully turbulent flow typically associated with chaos. The statistical
properties of the energy spectra in this regime in the strict semiclassical
limit obey the universal laws predicted by GOE of random matrix
theories (e.g. Bohigas 1991).
This has been studied in detail by Prosen and Robnik (1993,1994a,b) and by
Li and Robnik (1994c,1995b).
As for the curvature distribution it is
possibly statistically satisfactorily described by the random matrix theories
(Zakrzewski and Delande 1993, Delande and Zakrzewski 1994,
von Oppen 1994,1995), certainly in the tails of the curvature distribution,
but otherwise we have no reason to expect universality of the curvature
distribution for all values of the curvature.

\section{Level curvature distribution in fully chaotic systems}

In order to uncover the universal features in the curvature distribution
of the energy level flow, as predicted by Gaspard {\em et al} (1989,1990)
for the tails of the curvature distribution, we have to appropriately
normalize the curvature in such  a manner
that the normalized curvature is measured in certain natural units.
The answer is well known (Zakrzewski and Delande 1993,
Gaspard {\em et al} 1990), namely
\begin{equation}
k = \frac{K}{\pi \beta \rho <(dE_n/d\lambda)^2>},
\label{eq:curv}
\end{equation}
where $K=d^2 E_n/d\lambda^2$ is the actual curvature of the n-th energy level
with the eigenvalue $E_n$, i.e. its second derivative, $\beta$ is a constant
which according to the RMT is equal to 1,2 and 4 for GOE,
GUE and GSE, respectively, provided the classical dynamics of the system is
completely chaotic (ergodic).
In dynamical systems we use $\beta=1$ if the system has antiunitary symmetry,
$\beta=2$ if the system has no such symmetry and in case that the system has
half integer spin, an antiunitary symmetry but no rotational symmetry we take
$\beta=4$ (Berry and Robnik 1986, Robnik and Berry 1986, Robnik 1986).
 $\rho$ is the local density of energy levels,
which after unfolding is by construction equal to unity, and therefore the mean
value of all the derivatives of $E_n$ with respect to $\lambda$
vanish, in particular the average "velocity" is zero, i.e.
$<dE_n/d\lambda>=0$. $<(dE_n/d\lambda)^2>$
is the average of the squared "velocity" of the energy levels taken over a
suitable ensemble of consecutive energy levels (the spectral stretch). It can
be easily verified that $k$ thus defined is dependent on the parametrization
of the system and of its energy spectrum.
Namely, after re-parametrization $\mu=\mu(\lambda)$, we obtain
\begin{equation}
k_{\mu} = k_{\lambda} -\frac{v_{\lambda}}{\pi \beta \rho <v^2_{\lambda}>}
\frac{\mu^{\prime\prime}}{\mu^{\prime}},
\label{eq:newcurv}
\end{equation}
where $k_{\lambda}$, $v_{\lambda}$ are the (normalized) curvature and the
velocity calculated with respect to parameter $\lambda$, $k_{\mu}$ is the
(normalized) curvature in $\mu$-parametrization calculated according to
(\ref{eq:curv}), $\mu^{\prime\prime}$ and $\mu^{\prime}$ are the second and
first derivative of $\mu$ with respect to $\lambda$, whereas $\beta$ and
$\rho$  are as previously defined, and actually after unfolding $\rho=1$.
\\\\
The first immediate conclusion following this curvature
re-parametrization equation is that {\em as for the global curvature
distribution there cannot be any universality simply because the distribution
of the normalized curvature depends on parametrization}.
There are the following important remarks. The curvature distribution {\em is}
invariant with respect to the linear transformations $\mu=const\times\lambda$.
Knowing that typically the velocity distribution is Gaussian\footnote{We have
analyzed also the velocity distribution for many parameter values
$\lambda$ and in all cases the tails of the distribution are consistent with
the Gaussian distribution, and where the statistical significance is
sufficiently high we have also observed agreement for all velocities.}
 and thus rapidly
decaying with $v$ (Delande and Zakrzewski 1994), whilst the curvature
distribution typically has algebraic tails, we can see from equation
(\ref{eq:newcurv}) that {\em the tails} of the curvature distribution can be
universal, and they have been predicted for the first time by Gaspard {\em et
al} (1989,1990). This universal feature is correctly captured by the random
matrix model as conjectured by Zakrzewski and Delande (1993) and proven by von
Oppen (1994, 1995), namely
\begin{equation}
P(k) = \frac{C_{\beta}}{(1+k^2)^{(\beta+2)/2}},
\label{eq:vonOppen}
\end{equation}
where $\beta=1,2,4$ for GOE, GUE and GSE, respectively, and $C_{\beta}$ is the
normalization constant.
We should emphasize that universality at all $k$ cannot exist and therefore the
random matrix model (\ref{eq:vonOppen}) does not have the status of some
universal law, but is in fact just a case study, so that there
remains the open
question to clarify under what conditions it can be expected to model the
curvature distribution of a (one parameter) family of classically ergodic
Hamiltonian systems. There are cases, probably somewhat accidental, where the
agreement with RMT is very good at all $k$
and we shall report on such a case below.
\\\\
The first example that we give is the billiard system at $\lambda=0.41$, where
the classical dynamics is fully chaotic (probably ergodic, mixing and K) and we
have calculated the curvatures for 1,000 consecutive energy levels from
the 2,001-st to the
3,000-th eigenstate. The result is shown in figure 4b, where we
show the cumulative curvature distribution
\begin{equation}
 W(k) = \int_{-\infty}^{k} P(t) dt,
\label{eq:W(k)}
\end{equation}
in comparison with the GOE von Oppen formula based on (\ref{eq:vonOppen}).
For illustrative purposes we also show the histogram in figure 4a in comparison
with the von Oppen formula ($\beta=1$). The agreement is seen to be very good
for reasons that still have to be understood. Namely, there cannot be any
global universality and therefore there is no a priori reason that RMT model
should describe the dynamical systems' curvatures.
Interestingly, if we take the same
number of consecutive levels but at lower energies, namely from the 201-st to
the 1,200-th eigenstate, we observe a substantial degradation of the agreement
mainly manifested in the pronounced central peak at small $k$ as shown in
figures 5a,b.
Thus in this case in the semiclassical limit $E_n\rightarrow \infty$ the GOE
result seems to apply and the deviations from the von Oppen formula seem to be
attributed to low energies, i.e. not sufficiently small effective $\hbar$.
\\\\
We have checked the curvature distributions for at least one hundred values of
$\lambda$ covering the fully chaotic region at $\lambda\ge0.2791$ and the
transition region (KAM-like regime) of small $\lambda$. In all cases of full
chaos (ergodicity) the agreement with von Oppen formula was much worse than in
figure 4a,b even at high eigenstates and this is demonstrated collectively in
figure 6a,b where we show the curvature distributions for 34 values of
$\lambda$, at each of them taking the spectral stretch between the 2,001-st
and the 3,000-th eigenstate. The range of $\lambda$ is $[0.27,0.435]$ at equal
steps $\Delta \lambda=0.005$. The histogram in figure 6a gives an impression
that the agreement with GOE formula of von Oppen is very good, but the much
more informative cumulative plot in figure 6b reveals that the data are
statistically significantly settled (almost negligible fluctuations) and seem
to converge to a smooth distribution which, however, notably differs from GOE
distribution. It has to be verified whether this disagreement has a dynamical
reason and persists or even becomes stronger in the strict semiclassical limit,
or else it disappears in the strict semiclassical limit and by some
non-understood mechanism agrees with GOE. We think that this latter option is
unlikely, since universality in the large for curvature distribution does not
exist.
\\\\
As explained before the curvature distribution can have some universal features
and one of them is the algebraic tail predicted by Gaspard {\em et al}
(1989,1990)  and captured by von Oppen formula (\ref{eq:vonOppen}). We have
explicitly tested this aspect and the results are
shown in figure 7 for the systems of
figure 6a,b: The agreement as seen in the histogram of figure 7a seems to be
very good, however if we instead plot the data cumulatively, namely as a
plot of $\log(1-W(|k|))$ versus $\log(|k|)$, where we have no binning and
the numerical accuracy is fully respected, we must conclude that the agreement
is poor. In figure 8 we show the tail of the distribution in figure 4a,b and
as expected  the agreement with theory here is confirmed. Of course the very
last part of the tail is statistically not significant since there we have very
small number of objects and the expected statistical errors become huge.
One of the worst cases that we
found for the curvature distribution is at $\lambda=0.49$, which is deep in the
classically ergodic regime, for the eigenstates
2,001 through 5,000, where the distribution is extremely flat as shown in
figure 9a,b, but its tails nevertheless obey the theoretical prediction
satisfactorily as shown in figure 10, where in the range of $3\le|k|\le 10$ the
slope agrees with the RMT slope, namely -3 for $P(k)$ and -2 for $1-W(k)$.
As explained in the previous section the curvature distribution in figure 9a is
very flat paradoxically just due to the strong energy level repulsion which
gives rise to the laminar level flow (figures 1-3). Certainly, at higher
energies we would expect better agreement with the theoretical tails than
seen in figure 10a,b.

\section{Energy level curvature distribution in transition region between
integrability and chaos}

Now we turn to the problem of level curvatures of systems in the transition
region between integrability and full chaos, with mixed classical dynamics.
To our knowledge this problem has not been addressed so far in the literature.
But the problem is quite simple in the strict semiclassical limit. We adopt the
same assumptions that underly the Berry-Robnik approach (1984), especially the
validity of PUSC and the statistical independence of the energy level
subsequences associated with the regular components ($j=1$), and the
irregular components ($j=2,3,4,...,N$) (ordered in decreasing size).
These assumptions immediately imply that the curvature distribution $P(k)$ for
the entire spectrum is given by the simple additivity formula
\begin{equation}
P(k) = \sum_{j=1}^{N} \rho_j P_j(k),
\label{eq:ptotal}
\end{equation}
where $\rho_j$ is the fractional phase space volume of the given regular
($j=1$) components (lumped together in one subsequence),
or of the irregular components ($j=2,3,4,...,N$). Here
$P_1(k)$ is the curvature distribution of an integrable system and is just
a delta function $\delta (k)$, whereas $P_j(k)$ is the curvature
distribution for the spectral
subsequence associated with the $j$-th chaotic component, for which there is no
universal formula but under certain not yet understood conditions the von Oppen
formula of RMT might be a good model.
Its tail, however, must obey the universal
prediction of Gaspard {\em et al} (1989, 1990)
captured also in (\ref{eq:vonOppen}), in the strict semiclassical limit.
\\\\
In the almost integrable KAM-like regime at $\lambda=0.1$,
figure 11a,b, where the classical
 $\rho_1$ is estimated $\rho_1=0.88$ (Prosen and Robnik 1993), we find strong
delta spike with the smooth background modelled by $\rho_2 P_{{\rm GOE}}(k)$,
where
$\rho_2=1-\rho_1=0.12$ and $P_{{\rm GOE}}(k)$ is the GOE von Oppen formula
(\ref{eq:vonOppen}). Our judgement is that the agreement is satisfactory,
given the fact that we have only 1,000 objects, the curvatures of the
eigenstates from the 2,001-st through the 3,000-th.
Thus the agreement with two-component formula (\ref{eq:ptotal}) ($N=2$) is
reasonable.
As $\lambda$ is increased to $\lambda=0.15$ the area of the chaotic components
increases to $\rho_2=0.64$, $\rho_1=0.36$, we observe the drop of the strength
of the delta function as shown in figure 12a,b for relatively low states
(2,001-3,000). If we go higher in the semiclassical limit by considering the
eigenstates 5,001-7,000 we find considerable improvement of the agreement with
the semiclassical formula of von Oppen (\ref{eq:ptotal}), by using the
classical value $\rho_1=0.36$ for the dashed smooth background,
which  seems to be qualitatively well captured in figure 13a. Nevertheless,
the best fitting procedure to determine the parameter $\rho_1$ yields
$\rho_1=0.11$,  which is at variance with the classical value $\rho_1=0.36$.
By going higher in the semiclassical limit we
certainly expect further improvement, although the regime where
Berry-Robnik assumptions are satisfied is usually very hard to reach (Prosen
and Robnik 1993,1994a,b).
Therefore we should consider the results of figure 13a,b as the right trend
towards the semiclassical formula (\ref{eq:ptotal}).
This is also qualitatively confirmed in figure 14a,b where $\lambda=0.175$ and
the classical $\rho_1=0.17$.
\\\\
In our final plot we attempt to find the best fit of the semiclassical formula
(\ref{eq:ptotal}) with just two components, $j=1,2$, and one parameter
$\rho_1$,
to the numerical data (cumulative curvature distributions $W(k)$) for fifty
values of $\lambda$ covering the whole range of the mixed dynamics
$0\le\lambda\le 1/4$. We plot the quantity $D$ which is the
supremum norm of the difference between the numerical $W(k)$ and the best fit
$W(k)$. Unfortunately the significance of our semiclassical fits is not very
striking, but it is our impression that the behaviour has the right trend
towards Berry-Robnik surmise which underlies the semiclassical formula
(\ref{eq:ptotal}). One indication for this is the calculation of the curvatures
for $\lambda=0.15$ for twice higher energies, namely the eigenstates
5,001-7,000, where the agreement becomes indeed better and $D$ becomes almost
twice smaller (indicated by the arrow and the star symbol).
One should observe that the agreement is better at larger
$\lambda$'s and smaller $\rho_1$. In the range of classically almost or
completely ergodic motion $0.2791\le\lambda\le 1/2$ we have used not the
semiclassical formula (\ref{eq:ptotal}) but assumed instead $\rho_1=0$ and thus
used the von Oppen GOE formula (\ref{eq:vonOppen}) with $\beta=1$. Here for not
too large $\lambda$ the agreement is quite good and somewhere even excellent
like $\lambda=0.41$, as already shown in figure 4a,b and the tails in
figure 8a,b.
In fact the agreement is very good in the range of $0.27\le\lambda\le0.44$. For
reasons explained already in sections 2 and 3 for $\lambda$ close to 1/2 the
agreement with GOE becomes quite poor because of the flatness of the curvature
distribution which is predicted to gradually disappear when we go to higher
energies and thus to smaller effective $\hbar$, i.e. deeper in the
semiclassical limit. To demonstrate this we have calculated the curvatures at
$\lambda=0.49$ for twice larger energies in the range of eigenstates
4,001-5,000 resulting in the significantly smaller $D$ as indicated by the
arrow and by the star symbol.

\section{Discussion and conclusions}

We believe that in the present paper we provide further evidence that unlike
the spectral fluctuations and their statistics the curvature distribution
(defined by some local one parameter family of Hamiltonians) need not obey any
universal law such as e.g. the von Oppen formula (\ref{eq:vonOppen}) for the
random matrix models, even if the system is fully chaotic (ergodic), and even
if we are sufficiently far in the semiclassical limit. One
theoretical reason is the lack of the re-parametrization invariance of the
normalized curvature $k$, explicitly demonstrated by equation
(\ref{eq:newcurv}). On the other hand since the "velocities" (of the level
flow) typically are Gaussian distributed, the same equation shows that
asymptotically at sufficiently large $k$ we can have universality describing
the algebraic tail of the curvature distribution as predicted by
Gaspard {\em et al} (1989,1990) and
then correctly reproduced by von Oppen's formula
(\ref{eq:vonOppen}). Even being aware of these limitations regarding the
universality aspects of the curvature distribution we nevertheless find some
examples where the agreement with von Oppen's formula should be judged as very
good even for all $k$. It still  has to be understood under what conditions
precisely the RMT models would apply to dynamical systems. The
substantial (statistically highly significant) deviations from
GOE formula have been reported and emphasized already by Zakrzewski and Delande
(1993) for the case of the stadium and the
hydrogen atom in strong magnetic field and the nonuniversality aspects have
been discussed also by Takami and Hasegawa (1992,1994) although for relatively
low states and statistically still to be improved. Both groups proposed that
this is to be associated with the scarring and other localization phenomena
in the eigenstates. It is precisely these phenomena which imply the deviation
of spectral statistics from the Berry-Robnik surmise (1984) in the mixed
systems and manifest themselves in fractional power law level repulsion and
Brody-like behaviour as explained in detail in (Prosen and Robnik 1994a,b,
Li and Robnik 1995b). These phenomena disappear in the semiclassical limit
where we find uniformly extended chaotic states and the PUSC is fulfilled then,
and so is the Berry-Robnik (1984) surmise.
However, unlike the spectral statistics the curvature distribution need not
converge to the random matrix models not even in the strict semiclassical
limit where the localization phenomena disappear.
It is thus still an important open theoretical problem, especially in the
semiclassical level, to understand under what conditions the random matrix
models would apply to the curvature distribution of the quantal dynamical
systems. As explained in section 3, equation (\ref{eq:newcurv}), the global
universality of the curvature distribution could show up only in cases where
the natural parametrization is somehow restricted to the class of linear
transformations.
\\\\
The sensitivity of the eigenstates (eigenenergies and wavefunctions) on the
boundary data, of which one aspect is also the dependence of the eigenstates on
the billiard shape parameter, is an important theoretical problem.
If such sensitivity correlates with classical chaotic dynamics and at the same
time manifests itself in the accuracy of the purely quantal numerical methods,
then such a behaviour would be one important manifestation of quantum chaos.
We have recently demonstrated (Li and Robnik 1995d) that this is precisely the
case when applying the plane wave decomposition method of Heller (1991),
whilst e.g. in the boundary integral method such correlation does not show up
(Li and Robnik 1995c). Our present work where we analyze the curvature
distribution as a function of the billiard shape parameter is only
the first step in direction of this research.

\section*{Acknowledgments}

The financial support by the Ministry of Science
and Technology of the Republic of Slovenia is gratefully acknowledged.
We thank Dr. Vladimir Alkalaj, director of the National Supercomputer Center,
for the kind support in using the supercomputer facilities.

\vfill
\newpage
\section*{References}
B\"acker A, Steiner F and Stifter P 1994 {\em Preprint} DESY-94-213\\\\
B\"acker A 1995 {\em Diplomarbeit}, February 1995
Universit\"at Hamburg, II. Institut f\"ur
Theoretische Physik\\\\
Berry M V and Robnik M 1984 {\em J. Phys. A: Math. Gen.} {\bf 17} 2413\\\\
Berry M V and Robnik M 1986 {\em J. Phys. A: Math. Gen.} {\bf 19} 649\\\\
Bohigas O 1991  in {\em Chaos and Quantum Systems (Proc. NATO ASI Les Houches
Summer School)} eds M-J Giannoni, A Voros and J Zinn-Justin,
(Amsterdam: Elsevier) p87\\\\
Bruus H and Stone A D 1994 {\em Phys. Rev.} {\bf B50} 18275\\\\
Bruus H and Whelan N D 1995 {\em Preprint} Niels Bohr Institute, Copenhagen,
chao-dyn/9509005\\\\
Delande D and Zakrzewski J 1994 in {\em J. Phys. Soc. Japan} {\bf 63} Suppl A
101-121 (Proc. of QCTM) Hasegawa H (ed.)\\\\
Frisk H 1990 {\em Preprint} Nordita\\\\
Gaspard P, Rice S A and Nakamura K 1989 {\em Phys. Rev. Lett.} {\bf 63} 930\\\\
Gaspard P, Rice S A, Mikeska H J and Nakamura K 1990 {\em Phys. Rev.}
{\bf A42} 4015\\\\
Hayli A, Dumont T, Moulin-Ollagier J and Strelcyn J M 1987 {\em J. Phys. A:
Math. Gen.} {\bf 20} 3237\\\\
Heller E J 1991  in {\em Chaos and Quantum Systems (Proc. NATO ASI Les Houches
Summer School)} eds M-J Giannoni, A Voros and J Zinn-Justin,
(Amsterdam: Elsevier) p547\\\\
Li Baowen and Robnik M 1994a {\em J. Phys. A: Math. Gen.} {\bf 27} 5509\\\\
Li Baowen and Robnik M 1994b {\em to be published}\\\\
Li Baowen and Robnik M 1994c {\em Preprint CAMTP/94-11}, chao-dyn/9501022
unpublished but available upon request as a supplement to the paper Li and
Robnik 1995b\\\\
Li Baowen and Robnik M 1995a {\em J. Phys. A: Math. Gen.} {\bf 28} 2799\\\\
Li Baowen and Robnik M 1995b {\em J. Phys. A: Math. Gen.} {\bf 28} 4843\\\\
Li Baowen and Robnik M 1995c {\em Preprint} CAMTP/95-3, chao-dyn/9507002,
submitted to {\em J. Phys. A: Math. Gen.} in July\\\\
Li Baowen and Robnik M 1995d {\em Preprint} CAMTP/95-5, chao-dyn/9509016,
submitted to {\em J. Phys. A: Math. Gen.} in September\\\\
Markarian R 1993 {\em Nonlinearity} {\bf 6} 819\\\\
Percival I C 1973 {\em J. Phys. B: Atom. Molec. Phys.} {\bf 6} L229\\\\
Pomphrey N 1974 {\em J. Phys. B: Atom. Molec. Phys.} {\bf 7} 1909\\\\
Prosen T and Robnik M 1993 {\em J. Phys. A: Math. Gen.} {\bf 26} 2371\\\\
Prosen T and Robnik M 1994a  {\em J. Phys. A: Math. Gen.} {\bf 27} L459\\\\
Prosen T and Robnik M 1994b  {\em J. Phys. A: Math. Gen.} {\bf 27} 8059\\\\
Robnik M 1983 {\em J. Phys. A: Math. Gen.} {\bf 16} 3971\\\\
Robnik M 1984 {\em J. Phys. A: Math. Gen.} {\bf 17} 1049\\\\
Robnik M 1986 {\em Lecture Notes in Physics} {\bf 263} 120\\\\
Robnik M 1988 in {\em Atomic Spectra and Collisions in External Fields} eds. K
T Taylor, M H Nayfeh and C W Clark (New York: Plenum) 265-274\\\\
Robnik M 1993 {\em unpublished}\\\\
Robnik M and Berry M V 1986 {\em J. Phys. A: Math. Gen.} {\bf 19} 669\\\\
Stone A D and Bruus H 1993a {\em Physica } {\bf B189 } 43\\\\
Stone A D and Bruus H 1993b {\em Surface Sci.} {\bf 305} 490\\\\
Takami T and Hasegawa H 1992 {\em Phys. Rev. Lett.} {\bf 68} 419\\\\
Takami T and Hasegawa H 1994 in {\em J. Phys. Soc. Japan} {\bf 63} Suppl A
122-130 (Proc. of QCTM) Hasegawa H (ed.)\\\\
von Oppen F 1994 {\em Phys. Rev. Lett.} {\bf 73} 798\\\\
von Oppen F 1995 {\em Phys. Rev.} {\bf E51} 2647\\\\
Zakrzewski J and Delande D 1993 {\em Phys. Rev} {\bf E47} 1650\\\\

\newpage
\section*{Figure captions}
\bigskip
\bigskip

\noindent
{\bf Figure 1:} The unfolded energy level flow of the Robnik
billiard for the odd eigenstates from 1,000 to 1,050 (top).
The fine structures in the two small windows are magnified and plotted
in the bottom left and right boxes, respectively.
\bigskip
\bigskip

\noindent
{\bf Figure 2:} The same as figure 1 but for the eigenstates from 1,500 to
1,550.
\bigskip
\bigskip

\noindent
{\bf Figure 3:} The same as figure 1 but for the eigenstates from 2,000 to
2,050.
\bigskip
\bigskip

\noindent
{\bf Figure 4:} The histogram of the curvature distribution $P(k)$ (a)
and the cumulative curvature distribution $W(k)$ (b) for the Robnik billiard
at $\lambda=0.41$. The curvatures are calculated for 1,000
consecutive eigenstates, namely from 2,001 to 3,000.
The numerical results (solid line) are compared with the GOE distribution
of equation (\ref{eq:vonOppen}), with $\beta=1$ (dashed line). The agreement in
(b) seems to be surprisingly good.
\bigskip
\bigskip

\noindent
{\bf Figure 5:} The same as figure 4 but at lower eigenenergies, namely
from eigenstates 201 to 1,200. Please note the enhanced peak around $k=0$ in
curvature distribution $P(k)$ (a) and the deviation from GOE distribution in
$W(k)$ (b).
\bigskip
\bigskip

\noindent
{\bf Figure 6:} The curvature distribution for 34 values of $\lambda$ in the
range of $[0.27,0.435]$ at equal steps $\Delta \lambda=0.005$.
At each $\lambda$ the curvatures are calculated for the eigenstates
2,001 through 3,000. The $P(k)$ and $W(k)$ are shown in (a) and
(b), respectively. The numerical curve seems to be statistically settled (very
small statistical fluctuations giving rise to an apparently smooth step
function) but the deviation from the
GOE formula of equation (\ref{eq:vonOppen}) as seen in the
cumulative distribution $W(k)$ seems to be a physical effect.
\bigskip
\bigskip

\noindent
{\bf Figure 7:} The same data as in figure 6 but the $log(P(|k|)$ versus
$log(|k|)$ plot (a), and $log(1-W(|k|))$ versus $log(|k|)$ plot (b).
We show these plots in order to clearly display the behaviour in the tail of
the curvature distribution at large $|k|$.
\bigskip
\bigskip

\noindent
{\bf Figure 8:} The same data as in figure 4 but the $log(P(|k|)$ versus
$log(|k|)$ plot (a), and $log(1-W(|k|))$ versus $log(|k|)$ plot (b).
We show these plots in order to clearly display the behaviour in the tail of
the curvature distribution at large $|k|$.
\bigskip
\bigskip

\noindent
{\bf Figure 9:} The histogram of the curvature distribution $P(k)$ (a)
and the cumulative curvature distribution $W(k)$ (b) for $\lambda=0.49$.
The system is already completely chaotic but the deviation from GOE
distribution is very big. In this figure 3,000 consecutive eigenstates,
namely  2,001 to 5,000, are used to calculate the curvatures.
\bigskip
\bigskip

\noindent
{\bf Figure 10:} The same data as in figure 9 but the $log(P(|k|)$ versus
$log(|k|)$ plot (a), and $log(1-W(|k|))$ versus $log(|k|)$ plot (b).
We show these plots in order to clearly display the behaviour in the tail of
the curvature distribution at large $|k|$.
The slope of the tail seems to be in
better agreement with RMT model than the global $P(k)$ of figure 9.
\bigskip
\bigskip

\noindent
{\bf Figure 11:} The histogram of the curvature distribution $P(k)$ (a)
and the cumulative curvature distribution $W(k)$ (b) in transition region
at $\lambda=0.1$. In (a) we have the data in full line in comparison with
$\rho_2 P_{{\rm GOE}}(k)$ (dashed), where $\rho_2=0.12$ is determined by the
classical dynamics.  In (b) we plot the same data but cumulatively
in  comparison  with the theoretical cumulative distribution
$\rho_1\Theta(k) + \rho_2P_{{\rm GOE}}(k)$.
The data are the curvatures calculated for the eigenstates  2,001  through
3,000.
\bigskip
\bigskip

\noindent
{\bf Figure 12:} The same as figure 11 but for $\lambda=0.15$ and classical
$\rho_1=0.36$.
\bigskip
\bigskip

\noindent
{\bf Figure 13:} The same as figure 12 but the curvatures are taken from higher
eigenergies, namely from 5,001 through 7,000. The agreement with the
semiclassical formula in equation (\ref{eq:ptotal}) is definitely  better than
for the lower states of figure 12.
\bigskip
\bigskip

\noindent
{\bf Figure 14:} The same as figure 11 but for $\lambda=0.175$ and the
classical $\rho_1=0.17$.
\bigskip
\bigskip

\noindent
{\bf Figure 15:} The quantity $D$, defined as the supremum norm of the
difference of the numerical $W(k)$ and the theoretical $W(k)$,
versus $\lambda$. In the range of
$0<\lambda\le 1/4$, $D$ is calculated by comparison with the
the  best fit $W(k)$ of the semiclassical two-component formula of equation
(\ref{eq:ptotal}). For
$\lambda>1/4$, $D$ is calculated by comparison with
the GOE von Oppen formula (\ref{eq:vonOppen}). The $\Diamond$ is the result
for the energy stretch of eigenstates 2,001 through 3,000, while $\star$
at $\lambda=0.49$ represents the data from eigenstates 4,001 to 5,000 and
$\star$ at $\lambda=0.15$ represents the data from 5,001 to 7,000.

\end{document}